\documentclass{an}
\usepackage{graphicx}
\usepackage{times}
\newcommand{\ms}{m\ s\ensuremath{^{-1}}}
\newcommand{\mass}{M}

\newcommand{\mearth}{\ensuremath{\mass_\oplus}}

\begin{document}

\Pagespan{1}{}
\Yearpublication{2012}%
\Yearsubmission{2012}%
\Month{1}%
\Volume{333}%
\Issue{}%
\DOI{}%

\title{GJ~581 update: additional evidence for a Super-Earth in the habitable zone}

\author{Steven S. Vogt\inst{1}, R. Paul Butler\inst{2}, and Nader Haghighipour\inst{3}}
\titlerunning{GJ~581 update}
\authorrunning{S. S. Vogt, R. P. Butler, and N. Haghighipour}
\institute{UCO/Lick Observatory, Department of Astronomy and Astrophysics, University of California at Santa Cruz, Santa Cruz, CA 95064, USA \and
Department of Terrestrial Magnetism, Carnegie Institute of Washington, Washington, DC 20015, USA \and
Institute for Astronomy and NASA Astrobiology Institute, University of Hawaii-Manoa, Honolulu, HI 96822, USA}

\received{2012}
\accepted{2012}
\publonline{20 July 2012}


\abstract{We present an analysis of the significantly expanded HARPS 2011 radial velocity data set for GJ 581 that was presented by Forveille et al. (2011). Our analysis reaches substantially different conclusions regarding the evidence for a Super-Earth-mass planet in the star's Habitable Zone. We were able to reproduce their reported $\chi_{\nu}^2$ and RMS values only after removing some outliers from their models and refitting the trimmed down RV set. A suite of 4000 N-body simulations of their Keplerian model all resulted in unstable systems and revealed that their reported 3.6$\sigma$ detection of e=0.32 for the eccentricity of GJ~581e is manifestly incompatible with the system's dynamical stability. Furthermore, their Keplerian model, when integrated only over the time baseline of the observations, significantly increases the $\chi_{\nu}^2$ and demonstrates the need for including non-Keplerian orbital precession when modeling this system. We find that a four-planet model with all of the planets on circular or nearly circular orbits provides both an excellent self-consistent fit to their RV data and also results in a very stable configuration. The periodogram of the residuals to a 4-planet all-circular-orbit model reveals significant peaks that suggest one or more additional planets in this system. We conclude that the present 240-point HARPS data set, when analyzed in its entirety, and modeled with fully self-consistent stable orbits, by and of itself does offer significant support for a fifth signal in the data with a period near 32 days. This signal has a False Alarm Probability of $<4$\% and is consistent with a planet of minimum mass 2.2\mearth , orbiting squarely in the star's Habitable Zone at 0.13 AU, where liquid water on planetary surfaces is a distinct possibility.}

\maketitle


\section{Introduction and background}
\label{intro}

At a distance of only 20 light-years, the M3 dwarf GJ~581 has captured a special place in the public's perception of the rapidly growing tally of exoplanets. This is largely because the system is so nearby, and harbors at least four exoplanets,  two of which lie close to the formal boundaries of the star's classical liquid water Habitable Zone. Mayor et al. (\cite{M09}) (hereafter M09) summarizes the first four planets in this system, all discovered by the HARPS team. These four were subsequently confirmed by Vogt et al. (\cite{vogt10}) (hereafter V10) who combined the M09 HARPS data with an additional 122 HIRES measurements obtained over a much longer time baseline. From this combined data set, V10 reported evidence for an additional two planets, GJ 581f at 433 days, and GJ 581g at a period of 36.5 days. Soon after, Pepe et al. (\cite{pepe11}) reported that they had obtained an additional 60 HARPS measurements. Using only their set of 179 HARPS velocities, they were unable to confirm either planet f or g. This lack of confirmation was widely perceived to imply that, since the expanded HARPS data set, on its own, didn't see either planet f or g, neither could be there. Unfortunately, Pepe et al. (\cite{pepe11}) provided no new velocities beyond the 119 already available in M09. However, our own Monte Carlo simulations of such an expanded 179-point HARPS data set, having the same cadences and observing restrictions to which HARPS is subject, and the same susceptibility to aliases, indicated that is was unlikely that either planet f or g would have been significantly detectable in a 179-point HARPS data set alone.

Andrae et al. (\cite{and10}) criticized the V10 result on the grounds that its use of $\chi_{\nu}^2$ was not strictly valid in non-linear modeling situations. However the Andrae et al. (\cite{and10}) paper basically just made the point that, in non-linear situations, the actual value of the $\chi_{\nu}^2$ statistic can't be used to report a formal probability that a given model is correct. That was not done in any case in the V10 modeling of GJ 581. V10 thoroughly explored the $\chi_{\nu}^2$ surface, looking for the best global minima, and also exploring other less optimal minima for alternate acceptable solutions. V10 used Markov chain Monte Carlo methods and simulated annealing optimization to thoroughly explore solution spaces around all relevant $\chi_{\nu}^2$ minima and to quantify uncertainties on all model parameters. Despite the caveats raised by Andrae et al. (\cite{and10}), $\chi_{\nu}^2$ minimization certainly remains a completely valid method for optimization and for comparing solutions, though one must be aware of its uncertainties and its complex behavior.

Tuomi (\cite{tuomi11}) and Gregory (\cite{gregory11}) then published Bayesian analyses of both the combined and individual data sets of M09 and V10. The Tuomi (\cite{tuomi11}) study explicitly concluded that the eccentricities of all the known orbits in the GJ 581 system are consistent with zero, and also reported finding marginal evidence for the 433-day periodicity attributed to GJ 581f. But evidence for the 36-day period for GJ 581g did not clear Tuomi's adopted Bayesian Evidence Ratio detection threshold of 148:1. The Bayesian study of Gregory (\cite{gregory11}) similarly found the eccentricities for 3 of the 4 orbits in this system to be consistent with zero within the uncertainties, and also found evidence for a $\sim400$-day signal in the M09 HARPS data set alone.

\begin{table}
\caption{F11 HARPS radial velocities for GJ 581}\label{tab:rvdata_HARPS_GJ581}
\begin{tabular}{lll}
\hline \noalign{\smallskip}
JD & RV  & Uncertainty  \\
   & \multicolumn{2}{c}{(m\,s$^{-1}$)} \\
\noalign{\smallskip} \hline \noalign{\smallskip}
2453152.71289 & -10.25 & 1.10\\
2453158.66346 & -19.05 & 1.30\\
2453511.77334 &  -7.25 & 1.20\\
2453520.74475 &  10.35 & 1.40\\
2453573.51204 &   0.65 & 1.30\\
2453574.52233 &   9.05 & 1.10\\
2453575.48075 &   4.35 & 1.00\\
2453576.53605 &  -7.15 & 1.00\\
2453577.59260 & -10.85 & 1.20\\
2453578.51071 &   0.35 & 0.90\\
2453578.62960 &   2.25 & 1.10\\
2453579.46256 &  13.25 & 0.90\\
2453579.62105 &  14.75 & 1.10\\
2453585.46177 &   7.75 & 1.10\\
2453586.46516 &  -3.05 & 0.80\\
2453587.46470 & -17.25 & 1.60\\
2453588.53806 &  -8.15 & 2.60\\
2453589.46202 &   6.05 & 0.80\\
2453590.46390 &  12.75 & 0.80\\
2453591.46648 &   7.75 & 0.80\\
2453592.46481 &  -5.25 & 0.80\\
2453606.55168 &  14.85 & 2.10\\
2453607.50753 &  11.55 & 1.00\\
2453608.48264 &  -3.75 & 1.20\\
2453609.48845 & -11.35 & 1.60\\
2453757.87732 &   5.75 & 1.00\\
2453760.87548 &  -1.45 & 1.30\\
2453761.85922 &   7.45 & 1.40\\
2453811.84694 &   6.45 & 1.30\\
2453813.82702 &  -9.95 & 0.90\\
2453830.83696 &  -1.75 & 0.90\\
2453862.70144 &  -0.55 & 0.90\\
2453864.71366 &  13.85 & 1.10\\
2453867.75217 &  -9.25 & 1.10\\
2453870.69660 &   6.35 & 1.10\\
2453882.65776 & -11.35 & 0.90\\
2453887.69074 &  -5.65 & 0.80\\
2453918.62175 &   7.75 & 1.10\\
2453920.59495 & -18.45 & 1.00\\
2453945.54312 &   9.25 & 1.00\\
2453951.48593 &   6.55 & 0.80\\
2453975.47160 &  -7.35 & 1.00\\
2453979.54398 &  -7.35 & 1.30\\
2454166.87418 & -12.85 & 1.10\\
2454170.85396 &   7.95 & 0.90\\
2454194.87235 & -12.45 & 1.10\\
2454196.75038 &  16.35 & 1.20\\
2454197.84504 &  15.25 & 1.20\\
2454198.85551 &  -2.15 & 1.30\\
2454199.73287 &  -5.35 & 1.00\\
2454200.91092 &  -0.05 & 1.10\\
2454201.86855 &   9.35 & 1.00\\
2454202.88260 &  12.55 & 1.00\\
2454228.74156 &   8.55 & 1.10\\
2454229.70048 &  10.35 & 1.50\\
2454230.76214 &  -1.75 & 1.00\\
2454234.64592 &  14.65 & 1.20\\
2454253.63317 &  -9.35 & 1.00\\
\noalign{\smallskip}\hline
\end{tabular}
\end{table}
\setcounter{table}{0}
\begin{table}
\caption{(continued)}
\begin{tabular}{lll}
\hline \noalign{\smallskip}
JD & RV  & Uncertainty  \\
   & \multicolumn{2}{c}{(m\,s$^{-1}$)} \\
\noalign{\smallskip} \hline \noalign{\smallskip}
2454254.66481 &  -4.35 & 1.00\\
2454291.56885 &  -6.85 & 1.40\\
2454292.59081 &   0.25 & 0.90\\
2454293.62587 &   9.85 & 1.00\\
2454295.63945 & -10.35 & 1.10\\
2454296.60611 & -19.65 & 1.30\\
2454297.64194 &  -8.25 & 1.00\\
2454298.56760 &   7.75 & 1.10\\
2454299.62220 &  10.95 & 1.60\\
2454300.61911 &  -0.95 & 1.00\\
2454315.50749 &  14.05 & 1.70\\
2454317.48085 &  -6.95 & 1.00\\
2454319.49053 &   2.95 & 1.40\\
2454320.54407 &  10.95 & 1.00\\
2454323.50705 & -11.05 & 3.70\\
2454340.55578 &  -1.65 & 0.90\\
2454342.48620 &  19.05 & 1.10\\
2454349.51516 & -11.85 & 1.00\\
2454530.85566 &   8.25 & 1.00\\
2454550.83127 &   6.45 & 0.90\\
2454553.80372 & -12.85 & 0.80\\
2454563.83800 &  -3.95 & 0.90\\
2454566.76115 &   0.95 & 1.10\\
2454567.79167 &   9.35 & 1.00\\
2454569.79330 & -14.35 & 1.00\\
2454570.80425 & -14.35 & 1.00\\
2454571.81838 &   4.15 & 1.10\\
2454587.86197 &   3.55 & 1.60\\
2454588.83880 &   9.75 & 1.10\\
2454589.82749 &   6.05 & 1.10\\
2454590.81963 &  -4.95 & 1.00\\
2454591.81712 & -19.55 & 1.60\\
2454592.82734 &  -9.75 & 0.90\\
2454610.74293 &  19.55 & 1.10\\
2454611.71348 &   9.05 & 0.90\\
2454616.71303 &   8.75 & 1.40\\
2454639.68651 &  -9.55 & 1.10\\
2454640.65441 & -10.15 & 1.40\\
2454641.63171 &   1.65 & 1.00\\
2454643.64500 &   2.35 & 1.30\\
2454644.58703 &  -9.85 & 1.30\\
2454646.62536 &  -7.05 & 1.20\\
2454647.57912 &   7.85 & 1.10\\
2454648.48482 &  10.05 & 1.10\\
2454661.55371 & -11.65 & 1.20\\
2454662.54941 &  -1.05 & 1.40\\
2454663.54487 &  14.15 & 1.20\\
2454664.55304 &  13.25 & 1.30\\
2454665.56938 &   6.65 & 1.00\\
2454672.53172 &  -4.55 & 1.90\\
2454674.52412 &   9.35 & 1.40\\
2454677.50511 &  -9.55 & 1.10\\
2454678.55679 &   2.85 & 1.20\\
2454679.50403 &  13.25 & 1.70\\
2454681.51414 &   3.15 & 1.50\\
2454682.50334 &  -5.85 & 1.40\\
2454701.48507 &  14.15 & 1.30\\
2454703.51304 &  -1.75 & 1.30\\
\noalign{\smallskip}\hline
\end{tabular}
\end{table}
\setcounter{table}{0}
\begin{table}
\caption{(continued)}
\begin{tabular}{lll}
\hline \noalign{\smallskip}
JD & RV  & Uncertainty  \\
   & \multicolumn{2}{c}{(m\,s$^{-1}$)} \\
\noalign{\smallskip} \hline \noalign{\smallskip}
2454708.47905 &  -5.35 & 1.20\\
2454721.47303 & -14.45 & 1.30\\
2454722.47237 &   2.65 & 1.20\\
2454916.91735 &   4.55 & 1.00\\
2454919.77751 & -13.95 & 0.90\\
2454935.69136 & -13.75 & 1.10\\
2454938.77023 &   4.85 & 1.10\\
2454941.70399 & -12.45 & 1.00\\
2454946.74298 &  -4.95 & 1.10\\
2454955.79358 &  -6.75 & 1.00\\
2454989.67874 &  -5.55 & 1.00\\
2454993.61155 &  -6.45 & 1.00\\
2454998.65589 &   0.75 & 1.30\\
2455049.51551 &  -0.25 & 1.30\\
2455056.52501 &  12.45 & 2.90\\
2455227.84095 &  11.35 & 1.10\\
2455229.88062 &  -5.75 & 1.40\\
2455230.85894 &  -8.15 & 1.80\\
2455232.88302 &  16.25 & 1.80\\
2455272.83531 &  -2.85 & 1.10\\
2455275.80926 &   9.25 & 1.30\\
2455277.83041 &  -1.75 & 1.20\\
2455282.86587 &   1.45 & 1.20\\
2455292.84315 &   7.45 & 1.10\\
2455294.77402 & -17.25 & 1.00\\
2455295.68472 & -13.45 & 1.30\\
2455297.76797 &  10.05 & 1.00\\
2455298.73452 &   4.05 & 1.00\\
2455299.68212 &  -7.15 & 1.70\\
2455300.72869 & -15.15 & 1.30\\
2455301.84323 &  -6.25 & 1.20\\
2455305.80850 & -12.65 & 1.00\\
2455306.76724 &  -6.85 & 1.00\\
2455307.76067 &   8.65 & 0.80\\
2455308.75781 &  17.75 & 0.90\\
2455309.76544 &   4.45 & 0.90\\
2455321.70852 &  -5.65 & 1.00\\
2455325.66237 &   5.45 & 1.30\\
2455326.61457 &  -9.55 & 1.30\\
2455328.63743 &  -3.65 & 1.60\\
2455334.66359 &  14.05 & 1.30\\
2455336.78989 &   5.05 & 1.00\\
2455337.65473 &  -7.25 & 1.40\\
2455349.63634 &   3.85 & 3.00\\
2455353.57756 & -11.25 & 1.30\\
2455354.60681 & -16.35 & 1.30\\
2455355.53696 &  -1.15 & 1.30\\
2455359.56247 &  -7.75 & 1.20\\
2455370.57818 & -11.45 & 2.00\\
2455372.55366 &  12.65 & 1.50\\
2455373.60234 &  11.75 & 1.20\\
2455374.61617 &  -2.35 & 1.50\\
2455375.55663 &  -9.45 & 1.40\\
2455389.64756 &  11.95 & 1.50\\
2455390.54432 &   4.65 & 4.70\\
2455391.54670 & -13.25 & 1.40\\
2455396.49708 &  -8.45 & 2.10\\
2455399.54017 &  18.65 & 1.30\\
\noalign{\smallskip}\hline
\end{tabular}
\end{table}
\setcounter{table}{0}
\begin{table}
\caption{(continued)}
\begin{tabular}{lll}
\hline \noalign{\smallskip}
JD & RV  & Uncertainty  \\
   & \multicolumn{2}{c}{(m\,s$^{-1}$)} \\
\noalign{\smallskip} \hline \noalign{\smallskip}
2455401.52230 &  -0.25 & 1.70\\
2455407.49699 & -11.35 & 1.00\\
2455408.50168 &  -5.05 & 1.90\\
2455410.55603 &  16.85 & 1.20\\
2455411.51484 &   8.75 & 1.50\\
2455423.51171 & -11.05 & 1.10\\
2455427.49846 &   8.55 & 1.10\\
2455428.48093 &  -4.75 & 1.30\\
2455434.51127 & -15.65 & 1.20\\
2455435.48705 & -11.65 & 1.00\\
2455436.48340 &   3.65 & 0.90\\
2455437.51432 &  14.85 & 0.90\\
2455439.48708 &  -9.55 & 1.00\\
2455443.49986 &   2.95 & 2.10\\
2455444.48950 & -10.55 & 1.10\\
2455445.49328 & -21.95 & 1.20\\
2455450.48002 & -10.35 & 1.40\\
2455453.48660 &  14.85 & 1.20\\
2455454.47680 &   4.15 & 1.40\\
2455455.48896 &  -9.45 & 1.30\\
2455457.47397 &  -4.35 & 1.00\\
2455458.48996 &   6.65 & 1.30\\
2455464.48161 &  16.45 & 1.70\\
2455626.90847 &  -3.85 & 1.70\\
2455627.86994 & -16.65 & 1.20\\
2455629.88250 &   7.15 & 1.00\\
2455630.88945 &  14.55 & 1.20\\
2455633.83855 & -11.05 & 0.90\\
2455634.83780 &   1.45 & 1.10\\
2455635.80037 &  14.95 & 1.40\\
2455638.87580 & -13.95 & 1.10\\
2455639.82564 &  -8.45 & 1.00\\
2455641.85816 &   9.75 & 1.00\\
2455642.78865 &  -0.75 & 1.20\\
2455644.87268 &  -5.45 & 1.10\\
2455646.85119 &  12.95 & 1.20\\
2455647.86060 &   1.85 & 1.00\\
2455648.89760 &  -9.95 & 1.10\\
2455652.83978 &   3.75 & 1.10\\
2455653.72224 &  -8.65 & 1.10\\
2455654.68243 & -15.85 & 1.10\\
2455656.75878 &   6.15 & 1.00\\
2455657.76630 &  14.25 & 1.10\\
2455658.82034 &   4.65 & 1.50\\
2455662.76574 &  14.55 & 1.40\\
2455663.75875 &   7.25 & 3.00\\
2455672.71848 &   8.25 & 1.20\\
2455674.73187 &   3.35 & 1.00\\
2455675.77838 & -12.15 & 1.30\\
2455676.75948 &  -9.05 & 1.40\\
2455677.69188 &   3.75 & 1.50\\
2455678.76651 &  11.25 & 1.50\\
2455679.71364 &   6.45 & 0.90\\
2455680.62402 &  -2.15 & 1.30\\
2455681.67631 & -12.95 & 1.00\\
2455682.67267 &  -2.25 & 1.00\\
2455683.62142 &  13.45 & 1.00\\
2455684.67393 &  14.35 & 1.10\\
\noalign{\smallskip}\hline
\end{tabular}
\end{table}
\setcounter{table}{0}
\begin{table}
\caption{(continued)}
\begin{tabular}{lll}
\hline \noalign{\smallskip}
JD & RV  & Uncertainty  \\
   & \multicolumn{2}{c}{(m\,s$^{-1}$)} \\
\noalign{\smallskip} \hline \noalign{\smallskip}
2455685.64645 &   1.55 & 0.90\\
2455686.65353 &  -7.25 & 1.40\\
2455689.71145 &  12.55 & 1.30\\
2455690.73996 &  -1.15 & 1.60\\
2455691.69250 & -11.95 & 1.40\\
2455692.71193 & -12.75 & 1.20\\
2455693.75657 &  -2.25 & 1.00\\
2455695.62767 &  12.15 & 0.90\\
\noalign{\smallskip}\hline
\end{tabular}
\end{table}

At about the same time, Anglada-Escude \& Dawson (\cite{ang11}) (hereafter AD11) presented a detailed discussion of a particularly confusing situation with GJ 581 that arises from the first eccentricity harmonic of the 67-day planet GJ 581d. AD11 showed that any eccentricity of the 67-d orbit produces a harmonic signal near half that period of $\sim 33.5$ days. The period of GJ 581g reported by V10 is 36.56 days, and one of its yearly aliases occurs near a period of 1/p =  1/36.56  + 1/365.25, or $\sim 33.2$ days. Because this yearly alias of planet g lies close to the eccentricity harmonic of the 67-day planet d, AD11 suggested that the signal from planet g can be partially or even totally absorbed by the eccentricity of planet d. AD11 carried out statistical tests to quantify these interactions and calculated False Alarm Probabilities (FAP) of 0.11\% and 0.03\% for the signals associated with 581g and 581f respectively. They concluded that the presence of GJ 581g is well supported by the data presented by M09 and V10.

Clearly, additional high precision radial velocity is needed to confirm or reject the presence of either GJ 581f or g. The additional 60 HARPS measurements cited back in October 2010 by Pepe et al. (\cite{pepe11}), plus another full observing season of data were released in September 2011 by Forveille et al. (\cite{Forveille11}), (hereafter F11), bringing the total number of published HARPS velocities for GJ 581 to 240. The F11 release essentially doubled the amount of high precision HARPS data publicly available since M09. F11 then presented Keplerian models to that data set. Like Pepe et al. (\cite{pepe11}), they also chose to exclude all HIRES data from their analysis to avoid any risk of being misled by subtle low-level systematics in one dataset or the other. F11 presented two multi-planet Keplerian models to this HARPS-only data set. The first was a four-planet model with the eccentricities of all orbits allowed to float. We will hereafter refer to this as their Keplerian model. The second was a four-planet model with all-circular orbits. We will hereafter refer to this as their Circular model. Neither of these models incorporated mutual gravitational interactions between planets, which we will show is essential for this system. F11 then used their Keplerian model to assess the likelihood of the 36-day and 433-day planets GJ~581\,g and GJ~581\,f claimed by V10. F11 concluded that their four-planet Keplerian model's fit to their greatly expanded HARPS data set reveals no significant residual signals and thus that the HARPS data set contains no evidence for any planets beyond the four already claimed by M09.

Most recently, Tadeu dos Santos et al. (\cite{tds12}) (hereafter TDS12) presented a new analysis of the M09 HARPS and V10 HIRES data sets for GJ 581. In agreement with AD11, they conclude that the existence of the 36-day planet g is intimately related to the orbital elements of 67-day planet d, and that it is not possible to disconnect the existence of the former from the determination of the eccentricity of the latter. They do find evidence for the planet f signal, at a period of $\sim 455$ days, but with a confidence level of 4\%, essentially at their detection limit. As regards GJ 581g, they conclude that, from a statistical point of view, given the data sets of M09 and V10, it is not incorrect to state the existence of GJ 581g. However, this requires the assumption that all planets in this system are in essentially circular orbits, an assumption strongly supported by the above-mentioned Bayesian studies.

In this work, we present a re-analysis of the 240-point RV data set released by F11. We critically analyze their models, and the planetary system that they imply. In particular, we examine the dynamical stability of their Keplerian model. We then present our own gravitationally self-consistent four-planet model to the full 240-point HARPS data set which reaches substantially different conclusions than those of F11.

\section{Stability of the GJ~581 system}
\label{stabI}

We present, for easy reference in Table 1, the set of 240 HARPS velocities provided by F11. We also present for reference in Tables 2 and 3 respectively, the Keplerian and Circular models presented by F11. For our dynamical stability analyses, we simply adopt the parameters listed by F11.
Throughout, we assume a stellar mass of 0.31 $M_{\odot}$, though none of the
uncertainties take into account the uncertainty in the stellar mass.  For our
fitting we used relative HARPS RVs, i.e. we subtracted the mean of the RVs from
each RV.  In F11's announced fits, the planets were assumed to be on
non-interacting orbits. We can test their model systems' stability by choosing a series
of epochs which determine the initial starting mean anomalies (MA). We choose two
sets of 1000 initial epochs from 0 to 10,000 days from the epoch of the first RV
observation. In the first set, the starting epochs are evenly spaced in time,
and in the second set, the starting epochs are randomly chosen.

\begin{figure*}
\includegraphics[angle=-90,width=160mm]{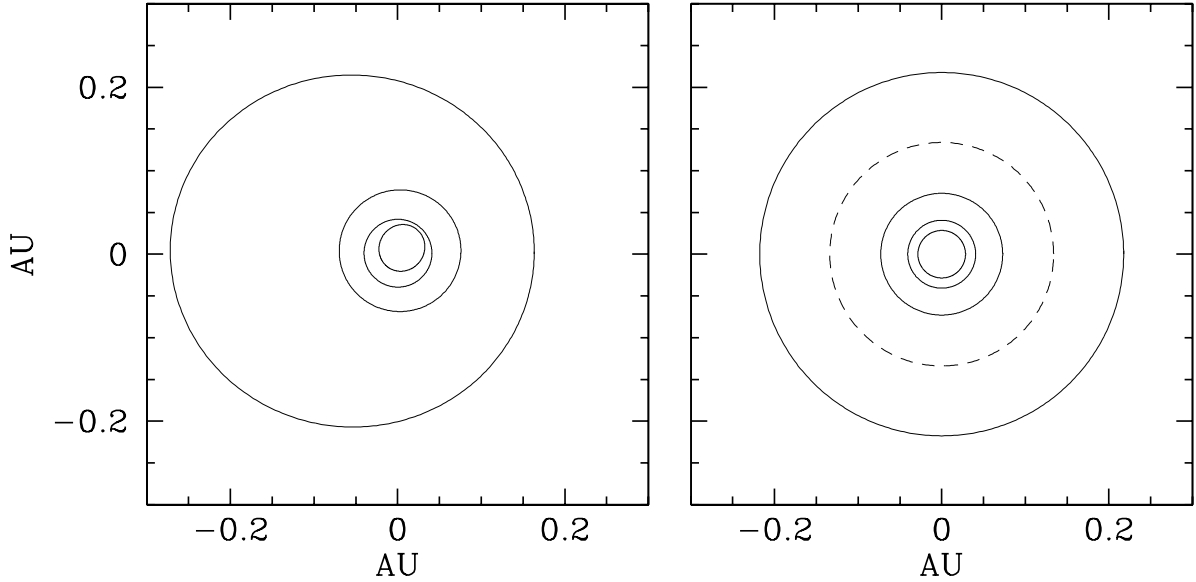}
\caption{Top view of the F11 Keplerian model \emph{(left panel)}, and their four-planet Circular model \emph{(right panel)}. The dashed orbit in the right panel marks the location of a potential fifth planet that will be discussed below.}
\label{topviews}
\end{figure*}

It is not clear if the parameters listed in F11 were Jacobi
or astrocentric elements.  We examined stability under each of these
two assumptions as well.  Thus, for each set of parameters given in
F11, we ran 4000 N-body simulations to test for stability.
For astrocentric elements, the positions and velocities of each planet are
relative to the star, and the mass in Kepler's third law is actually the sum of
the star and planet masses.  For Jacobi elements, the positions and velocities
of each planet are relative to the barycenter of the star and the masses with
smaller periods (the interior planets), and the mass in Kepler's third law is
actually the sum of the masses of the star, the interior planets, and the
(current) planet.  Under each assumption, the elements are then
straightforwardly calculated from the positions and velocities.

\begin{table*}
\caption{F11 Keplerian model}\label{F11_keplerian}
\begin{tabular}{lllll}
\hline \noalign{\smallskip}
Parameter & GJ~581\,e & GJ~581\,b & GJ~581\,c & GJ~581\,d  \\
\noalign{\smallskip} \hline \noalign{\smallskip}
$P$ (days)                  & 3.14945$\pm$0.00017 & 5.36865$\pm$0.00009 & 12.9182$\pm$0.0022 & 66.64$\pm$0.08 \\
$T0$ (JD-2400000)	 & 54750.31$\pm$0.13 & 54753.95$\pm$0.39 & 54763.0$\pm$1.6 & 54805.7$\pm$3.4 \\
$e$					& 0.32$\pm$0.09 & 0.031$\pm$0.014 & 0.07$\pm$0.06 & 0.25$\pm$0.09 \\
$\varpi$ ($^{\circ}$)   & 236$\pm$17 & 251$\pm$26 & 235$\pm$44 & 356$\pm$19 \\
$K$ (m\,s$^{-1}$)           & 1.96$\pm$0.20 & 12.65$\pm$0.18 & 3.18$\pm$0.18 & 2.16$\pm$0.22 \\
$m\,\sin{i}$ ($M_{\oplus}$) & 1.95 & 15.86 & 5.34 & 6.06 \\
$a$ (AU)                    & 0.028 & 0.041 & 0.073 & 0.22 \\
$N_{\rm meas}$              & \multicolumn{4}{c}{240}\\
$\chi_{\nu}^2$              & \multicolumn{4}{c}{2.57}\\
RMS (m\,s$^{-1}$)           & \multicolumn{4}{c}{1.79}\\
\noalign{\smallskip}\hline
\end{tabular}
\end{table*}

\begin{table*}
\caption{F11 circular model}\label{F11_Circular}
\begin{tabular}{lllll}
\hline \noalign{\smallskip}
Parameter & GJ~581\,e & GJ~581\,b & GJ~581\,c & GJ~581\,d  \\
\noalign{\smallskip} \hline \noalign{\smallskip}
$P$ (days)                  & 3.14941$\pm$0.00022 & 5.36864$\pm$0.00009 & 12.9171$\pm$0.0022 & 66.59$\pm$0.10 \\
$T$ (JD-2400000)	 & 54748.243$\pm$0.056 & 54750.199$\pm$0.012 & 54761.03$\pm$0.11 & 54806.8$\pm$1.0 \\
$K$ (m\,s$^{-1}$)           & 1.754$\pm$0.180 & 12.72$\pm$0.18 & 3.21$\pm$0.18 & 1.81$\pm$0.19 \\
$m\,\sin{i}$ ($M_{\oplus}$) & 1.84 & 15.96 & 5.41 & 5.26 \\
$a$ (AU)                    & 0.028 & 0.041 & 0.073 & 0.22 \\
$N_{\rm meas}$              & \multicolumn{4}{c}{240}\\
$\chi_{\nu}^2$              & \multicolumn{4}{c}{2.70}\\
RMS (m\,s$^{-1}$)           & \multicolumn{4}{c}{1.86}\\
\noalign{\smallskip}\hline
\end{tabular}
\end{table*}

Figure~1 shows top views of the GJ~581 system. The left panel shows the
Keplerian model of F11 (Table 2). The right panel shows their Circular model (Table 3) which,
as will be shown, is essentially identical to ours. The dashed orbit
denotes the position of a potential fifth planet in the system as will also be
discussed below. The close approach between the inner two orbits of the F11 Keplerian
model in the left panel of Figure 1 hints at possible dynamical instability. This was conclusively born out by detailed
N-body simulations as we will now describe.

Initial simulations of both of the F11 models already indicated that their eccentric configuration
quickly self-disrupts, while their circular configuration appears to be stable
on time scales $>10$ Myr.  Thus, all of our eccentric simulations were set up to
run up to 10 Myr. However, all of the simulations in which the planets were
started on (nearly) circular orbits were set up to run up to only 100,000 years. All simulations
for this work used a time step of 0.1 days and were done using the Hybrid simplectic integrator in
the Mercury integration package (Chambers \cite{mercury}), modified to include the
first order Post-Newtonian correction as in Lissauer \& Rivera (\cite{LR01}).

\begin{figure}
\includegraphics[angle=-90,width=80mm]{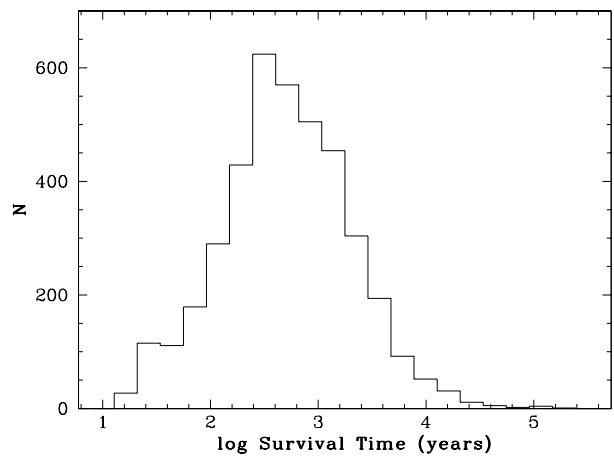}
\caption{Histogram of survival times for 4000 N-body simulations of the F11 Keplerian model.}
\label{survival}
\end{figure}

Table~4 lists the longest surviving simulation in each group of 1000 runs based
on the Keplerian fit from F11.  Among these 4000 simulations, not a single one
survived beyond 200,000 years, only seven survived to at least 50,000 years, and
only 24 survived for at least 20,000 years.  The shortest surviving systems
lasted only about 15 years.  Figure~2 shows a histogram of the survival times of
all 4000 N-body simulations of the four-planet Keplerian model of F11. Clearly,
the eccentric configuration presented in F11 is very unstable and therefore is
unphysical. All 4000 simulations ended with a collision between the inner two
planets.  This suggests that the eccentricity of the innermost planet plays a
significant role in system stability.  In contrast, we find that all 4000
simulations based on the circular fit are stable for at least 100,000 years.

\begin{table}
\caption{Longest survival times for each group of 1000 simulations based on the F11 Keplerian model}\label{sim_times}
\begin{tabular}{ll}
\hline \noalign{\smallskip}
Simulation group & Longest  \\
 &  surviving  \\
 &  simulation  \\
 &  (yrs) \\
\noalign{\smallskip} \hline \noalign{\smallskip}
astrocentric elements, evenly spaced epochs & 128300\\
astrocentric elements, randomly spaced epochs & 122900\\
Jacobi elements, evenly spaced epochs & 132400\\
Jacobi elements, randomly spaced epochs & 198900\\
\noalign{\smallskip}\hline
\end{tabular}
\end{table}

\section{New circular fits to the HARPS RVs}
\label{newcircfits}

In this section we present our own series of fits to the HARPS RVs of F11.  All fits in
this work were done using the Levenberg-Marquardt (LM) algorithm
(Sect.~15.5 in Press et al. \cite{Press07}), and correspond to epoch JD~2453152.712.
For those fits presented here that are intended to reproduce the F11 result, we do
not model the mutual interactions between the planets.

To determine uncertainties, we use the bootstrap method (Sect.~15.6 of
Press et al. \cite{Press07}).  We generate 1000 bootstrap RV sets, and we fit each of these
sets using our best-fit parameters in the initial guesses.  The uncertainties
in the orbital parameters are just the standard deviations of the fitted
parameters for the bootstrapped RV sets.

Following Gilliland \& Baliunas (\cite{GB87}), we also show the error-weighted Lomb-Scargle (wLS)
periodograms of the actual RV set as well as the residual RVs after fitting one,
two, three, and four planets.  We carry out a Monte Carlo false alarm probability (MCFAP)
analysis in which we use not only the actual RV set but also 1000 sets of mock
RVs for which we use the times of observations presented in F11, but
scramble the observed or residual RV values.  We define the MCFAP as the
fraction of the periodograms of the bootstrapped RVs or residuals that show a
peak that is at least as tall as the peak in the real RVs or residuals.

In addition to MCFAPs, we also give F-test probabilities for our fits.  We use
two methods to calculate the F-test statistic. The first method is based on
the difference in the RMS of fits (Sect.~14.2 in Press et al. \cite{Press07}).  The second is
based on the difference in reduced chi-squared ($\chi_{\nu}^2$) (Chapter 13 in
Frieden \cite{Frieden01}).  Small probabilities indicate statistically significant
differences.  For reference, we refer to the first and second F-test
probabilities as FT$_{\rm RMS}$ and FT$_{\chi2}$, respectively.

Some of the analysis here is based on the SYSTEMIC Console (Meschiari et al. \cite{Meschiari09}, \cite{Meschiari11}).
However, the results are based on fits done with a separate code in which the fitted
elements are astrocentric. Since the fitting in the Console is done explicitly in Jacobi elements,
in a few cases, these elements were then converted into Jacobi elements. Note that, for
multi-planet systems, conversion between the two coordinate systems results in mapping
circular orbits into slightly eccentric orbits (except for the innermost planet).  As a result,
there will be some discrepancies between the results given in this work and results obtained
with the SYSTEMIC Console. Discrepancies also arise from differences in the details in
implementing the LM algorithm in the code used for this work and in the Console. One notable
implementation difference is that, in the Console, $\chi_{\nu}^2$ values are based on assuming that five
parameters are added per planet while in the code used for this work, $\chi_{\nu}^2$ values are based
on the actual number of parameters that are allowed to vary. Although differences exist, the
almost identical values in the residual RV root-mean-square (RMS) values in corresponding
fits indicate that we are actually obtaining statistically identical fits.

\begin{figure}
\includegraphics[angle=-90,width=80mm]{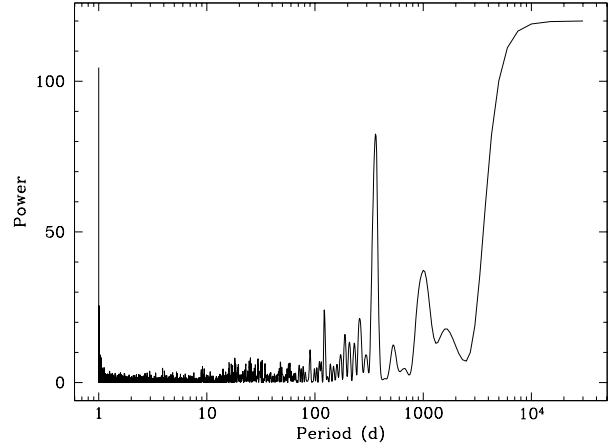}
\caption{Power spectral window of the HARPS RVs for GJ~581.}
\label{PSW}
\end{figure}

Figure~\ref{PSW} shows the power spectral window (PSW) of the HARPS RVs.  The
four most prominent and relevant periodicities occur at 354 days (roughly 1
year), 1 day, 978.5 days, and 122.4 days (roughly 4 months).  Very significantly,
there is no strong periodicity near the lunar synodic period.  We find
that the lunar synodic period and near-integer and sometimes even half-integer
multiples of this period can result in significant confusion in the detection
of real, small amplitude signals that are near these periodicities.  This is a
hardship related to being constrained to observing mostly in lunar bright time.
These alias issues, e.g. Dawson \& Fabrycky (\cite{DF10}), can gradually be removed as
the length of the observation baseline increases and the potential
real and false signals are observed at different phases.  The periodicities at larger
multiples of the lunar synodic period take longer to be removed.  Possibly, the
122-day periodicity is a remnant of the effect of the lunar synodic period in
the HARPS RVs. However, the longer periodicities in the PSW are somewhat easy to
identify as associated with prolonged stretches when the star was not observed.

We first analyze the system(s) assuming that the orbits are non-interacting
astrocentric circles (since we are attempting to reproduce the non-interacting
F11 models).  A constant RV model has $\chi_{\nu}^2=74.2672$ and RMS=9.9113 m\,s$^{-1}$.

Figure~4 shows periodograms of the data (top panel) and of the residuals for the
one-, two-, and three-planet fits (successive descending panels) from our
model using non-interacting circular orbits. The top panel
shows the dominant period in the system, a strong peak near 5.4 days. Three MCFAP
levels of 0.1, 1, and 10\% are shown. The 5.4-day signal has a MCFAP\,$\ll0.001$.
We fit a sinusoid with period ($P$) 5.3687 days and semi-amplitude ($K$)
12.9988 m\,s$^{-1}$. With the assumed stellar mass of 0.31\,$M_{\odot}$, this
corresponds to a minimum mass ($m\,\sin{i}$) of 16.30\,$M_{\oplus}$.  Our two F-test
values for this first planet are FT$_{\rm RMS}=3.5\times10^{-49}$ and
FT$_{\chi2}=1.5\times10^{-107}$.

\begin{figure}
\includegraphics[angle=0,width=80mm]{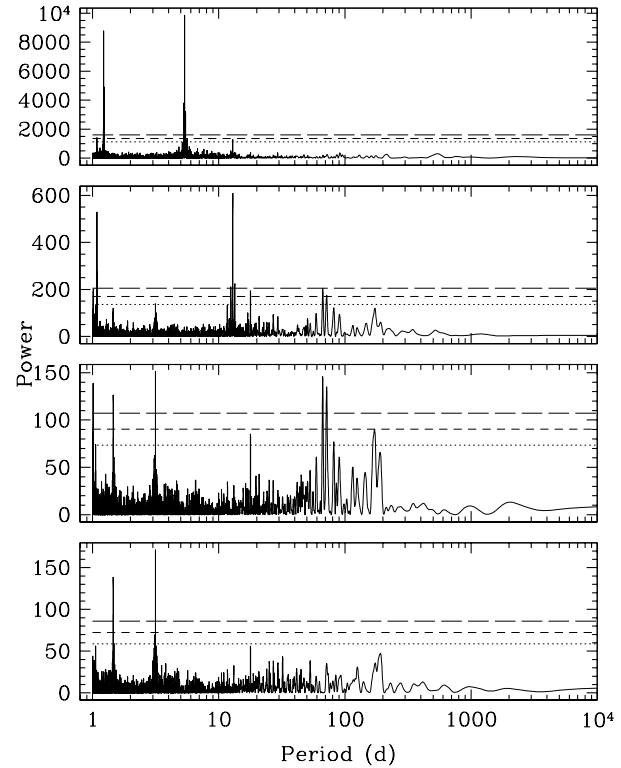}
\caption{Successive wLS periodograms of the fit residuals for GJ~581 using non-interacting circular orbits. The models are listed in order of 0, 1 ,2, and 3 planets from top to bottom. Three MCFAP levels of 0.1, 1, and 10\% are shown.}
\label{periodograms}
\end{figure}

The second panel of Figure~4 shows the wLS periodogram of the residuals of the
one-planet fit.  The peak at $\sim12.9$ days also has a MCFAP$\,\ll0.001$.  An
all-circular two-planet fit achieves $\chi_{\nu}^2=4.9666$ and
RMS=2.7086 m\,s$^{-1}$. The fitted astrocentric $P$s, $K$s, and
$m\,\sin{i}$\,s are 5.3686 and 12.9287 days, 12.6374 and 3.3049 m\,s$^{-1}$,
and 15.85 and 5.56 $M_{\oplus}$, respectively.  Our F-test values for the second
planet are FT$_{\rm RMS}=2.2\times10^{-5}$ and FT$_{\chi2}=7.8\times10^{-31}$.

\begin{table*}[tbh]
\caption{Astrocentric, circular, non-interacting orbital model}
\label{4pl_circ_nint}
\begin{tabular}{lllll}
\hline \noalign{\smallskip}
Parameter & GJ~581\,e & GJ~581\,b & GJ~581\,c & GJ~581\,d  \\
\noalign{\smallskip} \hline \noalign{\smallskip}
$P$ (days)                  & 3.1494$\pm$0.0263 & 5.3694$\pm$0.0135 & 12.934$\pm$0.125 & 66.71$\pm$3.67 \\
$a$ (AU)                    & 0.028459$\pm$0.000165 & 0.0406162$\pm$0.0000677 & 0.072983$\pm$0.000471 & 0.2179$\pm$0.0106 \\
$K$ (m\,s$^{-1}$)           & 1.749$\pm$0.384 & 12.74$\pm$1.06 & 3.212$\pm$0.517 & 1.806$\pm$0.379 \\
$m\,\sin{i}$ ($M_{\oplus}$) & 1.836$\pm$0.404 & 15.98$\pm$1.32 & 5.400$\pm$0.869 & 5.25$\pm$1.11 \\
MA ($^{\circ}$)              & 138.5$\pm$40.3 & 338.9$\pm$16.4 & 175.2$\pm$48.7 & 235.8$\pm$37.6 \\ \hline
fit epoch (JD)              & \multicolumn{4}{c}{2453152.712}\\
$N_{\rm meas}$              & \multicolumn{4}{c}{240}\\
$\chi_{\nu}^2$              & \multicolumn{4}{c}{2.881}\\
RMS (m\,s$^{-1}$)           & \multicolumn{4}{c}{2.011}\\
\noalign{\smallskip}\hline
\end{tabular}
\end{table*}

\begin{table*}[tbh]
\caption{Astrocentric, circular, non-interacting orbital model using the trimmed RV set that nearly reproduces the RMS and $\chi_{\nu}^2$ in F11}
\label{removed5_4pl_circ_nint}
\begin{tabular}{lllll}
\hline \noalign{\smallskip}
Parameter & GJ~581\,e & GJ~581\,b & GJ~581\,c & GJ~581\,d  \\
\noalign{\smallskip} \hline \noalign{\smallskip}
$P$ (days)                  & 3.1494$\pm$0.0640 & 5.3694$\pm$0.0153 & 12.9333$\pm$0.0341 & 66.69$\pm$4.38 \\
$a$ (AU)                    & 0.028459$\pm$0.000415 & 0.0406162$\pm$0.0000765 & 0.072981$\pm$0.000128 & 0.21783$\pm$0.00816 \\
$K$ (m\,s$^{-1}$)           & 1.767$\pm$0.402 & 12.766$\pm$0.952 & 3.283$\pm$0.592 & 1.724$\pm$0.362 \\
$m\,\sin{i}$ ($M_{\oplus}$) & 1.855$\pm$0.423 & 16.01$\pm$1.19 & 5.52$\pm$1.00 & 5.01$\pm$1.05 \\
MA ($^{\circ}$)              & 141.3$\pm$40.5 & 339.1$\pm$14.2 & 171.9$\pm$53.2 & 231.6$\pm$40.2 \\ \hline
fit epoch (JD)              & \multicolumn{4}{c}{2453152.712}\\
$N_{\rm meas}$              & \multicolumn{4}{c}{235}\\
$\chi_{\nu}^2$              & \multicolumn{4}{c}{2.686}\\
RMS (m\,s$^{-1}$)           & \multicolumn{4}{c}{1.857}\\
\noalign{\smallskip}\hline
\end{tabular}
\end{table*}

The third panel down in Figure~4 shows the wLS periodogram of the residuals of
the two-planet fit. Both peaks at $\sim3.15$ and $\sim66.7$ days have
FAP$<0.001$. Since they are of approximately the same power, it is not
immediately clear which to fit next.  However, in the four-planet fit to be
presented below, the $K$ for the $\sim66.7$-day planet is larger than that of
the $\sim3.15$-day planet.  For this reason, we take the ``third'' planet to be
at $\sim66.7$ days.  Note that choosing either peak and carrying out a circular
three-planet fit leaves the other peak in the residuals periodogram and that
the two resulting four-planet fits are statistically identical.  An all-circular
three-planet fit achieves $\chi_{\nu}^2=4.0204$ and RMS=2.3954 m\,s$^{-1}$.  The
fitted astrocentric $P$s, $K$s, and $m\,\sin{i}$\,s are 5.3687, 12.9306, and
66.8169 days, 12.7011, 3.1482, and 1.6578 m\,s$^{-1}$, and 15.93, 5.29, and
4.82 $M_{\oplus}$, respectively.  Our F-test values for the third planet are
FT$_{\rm RMS}=0.058$ and FT$_{\chi2}=6.9\times10^{-11}$.

The bottom panel in Figure~4 shows the wLS periodogram of the residuals of the
three-planet fit. The peak at $\sim3.15$ days has MCFAP$<0.001$. Fitting that out
results in a four-planet all-circular fit that achieves a $\chi_{\nu}^2=2.8806$
and RMS=2.0107 m\,s$^{-1}$. Our F-test values for the fourth planet are
FT$_{\rm RMS}=0.0068$ and FT$_{\chi2}=1.1\times10^{-16}$. The implied stellar
jitter for this fit (i.e. the value of the stellar jitter that is required to make
$\chi_{\nu}^2=1.0$) is 1.52 \ms.

Table~5 lists our best-fit astrocentric parameters under the assumption that the
planets do not interact. We list astrocentric parameters here because the
conversion to Jacobi parameters results in non-zero eccentricities for the outer
three planets (with the largest value of $\sim2.6\times10^{-4}$). For all circular orbits, the mean anomalies (MA) are defined relative to
the periastron longitude, which is assumed to be zero (and in the sky).

Comparing our Table~5 with the Circular model from F11 (provided for reference here in Table 3 above) shows some
notable differences. First, we simply used the parameters from their Table 2 as
an initial guess. This results in $\chi_{\nu}^2$ and RMS values closer to our
values rather than to the values of 2.70 and 1.86 \ms\ reported by F11. We
used the simulating annealing algorithm in the SYSTEMIC Console to see
if we could obtain a fit with RMS as low as that in F11 but could not get an
RMS below 2.0105 m\,s$^{-1}$.  We then tried successively removing
those observations with the largest reported uncertainties, and re-fitting the four-planet
configuration above. Again, we were unable to reproduce the
RMS and $\chi_{\nu}^2$ of F11. Our best fit to this trimmed (235-point) RV
set achieved $\chi_{\nu}^2=2.9192$ and RMS=1.9620 m\,s$^{-1}$.

We next instead tried gradually removing, one at a time, points with the largest
residual RVs from the model and then re-fitting the four-planet configuration. It was only when
we had removed the five points with the largest residual RVs from the
four-planet fit that we were able to get RMS and $\chi_{\nu}^2$ values very near
the values reported by F11. Table 6 shows that model, computed using 235 of the 240 HARPS velocities.

All five of the omitted velocities
have a residual RV (from the model) $\geq 5$ m\,s$^{-1}$ whereas the next largest residual RV
is 4.82 m\,s$^{-1}$. As an additional check, we repeated this procedure using both the F11 Keplerian and Circular
models as fit to all 240 velocities, with similar results: the five worst-fitting points from the F11 models in each case had to be removed to recover the $\chi_{\nu}^2$ and RMS values reported by F11. Table 7 shows our rank-ordered top ten residuals from the F11 Keplerian model and the resulting RMS and $\chi_{\nu}^2$ of the F11 model when all points up to and including that point have been removed. Here, we used the SYSTEMIC console, working in Jacobi coordinates. In each case, the Mean Anomalies and velocity zero point were allowed to re-optimize. The RMS and $\chi_{\nu}^2$ values underlined in bold correspond to the values reported in Table 2 of F11. Here, we find that the F11 results are recovered precisely only when the top 5 points with the largest residuals to the F11 model are removed.

\begin{table}
\caption{Effect of removing points based on residuals from the F11 Keplerian model}
\label{F11_kepleriantrim}
\begin{tabular}{lllll}
\hline \noalign{\smallskip}
JD & Residual & $N_{\rm obs}$ & RMS &  $\chi_{\nu}^2$  \\
\noalign{\smallskip} \hline \noalign{\smallskip}
            &       & 240 & 1.9601 & 2.7438 \\
2455349.64  & 6.59  & 239 & 1.9178 & 2.7342 \\
2454672.53  & 5.69  & 238 & 1.8872 & 2.7059 \\
2455295.68 & -5.63  & 237 & 1.8538 & 2.6316 \\
2455390.54  & 5.28  & 236 & 1.8255 & 2.6379 \\
2454678.56  & 5.19  & 235 & {\bf \underbar{1.7999}} & {\bf \underbar{2.5616}} \\
2455408.50  & 5.05  & 234 & 1.7735 & 2.5396 \\
2453761.86 & -4.92  & 233 & 1.7474 & 2.4951 \\
2454935.69 & -4.84  & 232 & 1.7200 & 2.4135 \\
2455282.87 & -4.67  & 231 & 1.6944 & 2.3526 \\
2454989.68  & 4.36  & 230 & 1.6757 & 2.2730 \\
\noalign{\smallskip}\hline
\end{tabular}
\end{table}

We repeated this analysis using the F11 Circular model. Again, we allowed the velocity zero point and Mean Anomalies to re-optimize for each successive case. The results are listed in Table 8. In this case, again we find that the top 5 points with the largest residuals from the F11 Circular fit had to be discarded to recover the RMS and $\chi_{\nu}^2$ values reported in Table 2 of F11. However, the points are slightly different than the ones required for the F11 Keplerian model. Most, but not all of these apparently omitted points agreed across all 3 model analyses. But since there was not exact agreement on which and how many were omitted across all three analyses, we cannot say with 100\% certainty exactly which points appear to have been omitted from the F11 analysis.

\begin{table}
\caption{Effect of removing points based on residuals from the F11 Circular model}
\label{F11_circular}
\begin{tabular}{lllll}
\hline \noalign{\smallskip}
JD & Residual & $N_{\rm obs}$ & RMS &  $\chi_{\nu}^2$  \\
\noalign{\smallskip} \hline \noalign{\smallskip}
           &        & 240 & 2.0080 & 2.8806 \\
2454672.53  & 6.30  & 239 & 1.9712 & 2.8445 \\
2455408.50  & 5.80  & 238 & 1.9395 & 2.8151 \\
2455295.68 & -5.72  & 237 & 1.9062 & 2.7414 \\
2455349.64  & 5.35  & 236 & 1.8789 & 2.7396 \\
2453761.86 & -4.92  & 235 & {\bf \underbar{1.8530}} & {\bf \underbar{2.6941}} \\
2455370.58  & 4.85  & 234 & 1.8303 & 2.6799 \\
\noalign{\smallskip}\hline
\end{tabular}
\end{table}

F11 specifically drew attention to points with the largest fit residuals, stating ``The largest
residuals such as those which stand out at phases 0.5 to 0.6 in the
Gl 581b panel of Figure 1, correspond to spectra with low S/N
ratio (under 35, compared to a median of 46), obtained through
either clouds or degraded seeing. Ignoring those measurements
produces visually more pleasing figures, but leaves the orbital
parameters essentially unchanged and only modestly lowers the
$\chi_{\nu}^2$ of the least square fit. We chose to retain them, for
the sake of simplicity.''

Our failure to reconcile their reported RMS and $\chi_{\nu}^2$ values obliges us to conclude that some unspecified number of points were in fact omitted by F11 when computing the $\chi_{\nu}^2$ and RMS for their models. Most of the apparently omitted points were not distinguishable on the basis of excess uncertainty due to low S/N, clouds, or degraded seeing. Rather, it took removal of those 5-6 points with the largest deviation from either of the F11 models in order to be able to reproduce their RMS and $\chi_{\nu}^2$ values. F11 state that they chose to retain all points, for the sake of simplicity. However, while these points may be present in their phased plots, they do not seem to have been included in their RMS and $\chi_{\nu}^2$ calculations. More troublingly, they also do not appear to be present in the calculations underlying their residuals periodograms. We similarly examined the RVs and fit from M09, and found that, again, the five observations with the largest residual RVs (there with residuals $\geq 3.5$ m\,s$^{-1}$) to our nominal four-planet fit had to be omitted in order to accurately reproduce their RMS and $\chi_{\nu}^2$ values (they actually give $\sqrt{\chi_{\nu}^2}$ values in that work).

We also draw attention to the fact that, in both our Tables~5 and 6, our bootstrap uncertainties are significantly larger than
the uncertainties listed in F11. Below, we will show that there is somewhat better agreement
when we include a fifth planet. These are very small amplitude signals however,
and the bootstrap method is effectively removing random points. With such small
amplitude signals, it may not take the removal of too many points to obtain
significantly different fits. It is not clear from the F11 paper how they
computed their uncertainties.

Figure~5 compares the 4-planet residuals periodograms to the F11 Circular model, both with and without discarding of points. MCFAP levels of 0.1, 1, and 10\% are shown (top to bottom). The top panel of Figure~5 shows the periodogram of the residuals of the F11 four-planet all-circular fit, done without including the dropped points discussed above. The bottom panel of Figure 5 shows the periodogram of the residuals to our astrocentric four-planet all-circular model (both interacting or non-interacting) which includes all 240 HARPS velocities. Not unexpectedly, the power of both the 32-day and 190-day residuals peaks (top panel) is significantly reduced by the omission of the five worst-fit outlier points. Any omission of points based on deviation from a given model will unfairly lessen the power remaining from any additional signals in the data not represented by that model.

As will be
discussed below, the corresponding residuals periodogram for the fully
self-consistent (interacting) four-planet model is essentially identical to
this (lower) plot in Figure~5. The top panel residuals periodogram in Figure~5 was not
shown or discussed by F11 even though they had presented essentially that same model.
Both residual periodograms of Figure~5  show clear peaks near 32 and 190 days.
The peak at $\sim32.1$ days has a MCFAP of 2.9\% for the bottom panel and 11.9\% for the top panel.
So, omitting the five worst-fitting outliers effectively quadrupled the FAP for the 32-day signal, from 2.9\% to 11.9\%, thereby significantly and unfairly weakening the case for any further planets in this system. An all-circular five-planet
fit using this 32.1-day period for the 5th planet achieves $\chi_{\nu}^2=2.5701$ and RMS=1.9067 m\,s$^{-1}$.  F-test values for the fifth planet are FT$_{\rm RMS}=0.4156$ and FT$_{\chi2}=5.5\times10^{-6}$.

Table~9 lists our best-fit astrocentric parameters under the assumption that the
five planets do not interact. Noting the similarity of the period of the 32-day to the 36.6-day period of GJ~581g,
we retain that nomenclature here to avoid confusion from renaming planet-f. For the four planets they have in
common, the uncertainties for our five-planet fit listed in Table~9 are now in
somewhat better agreement with the uncertainties listed in the bottom half of
Table~2 in F11 (and again provided for reference in Table 3 above). The relatively larger uncertainties for the
parameters of the fifth planet, indicative of poorly-determined parameters, are consistent with
the relatively large MCFAP and FT values. So while this 5th potential signal is interesting, its MCFAP and FT values do not yet meet our formal criteria to qualify as a firm detection.

\begin{figure}
\includegraphics[angle=0,width=80mm]{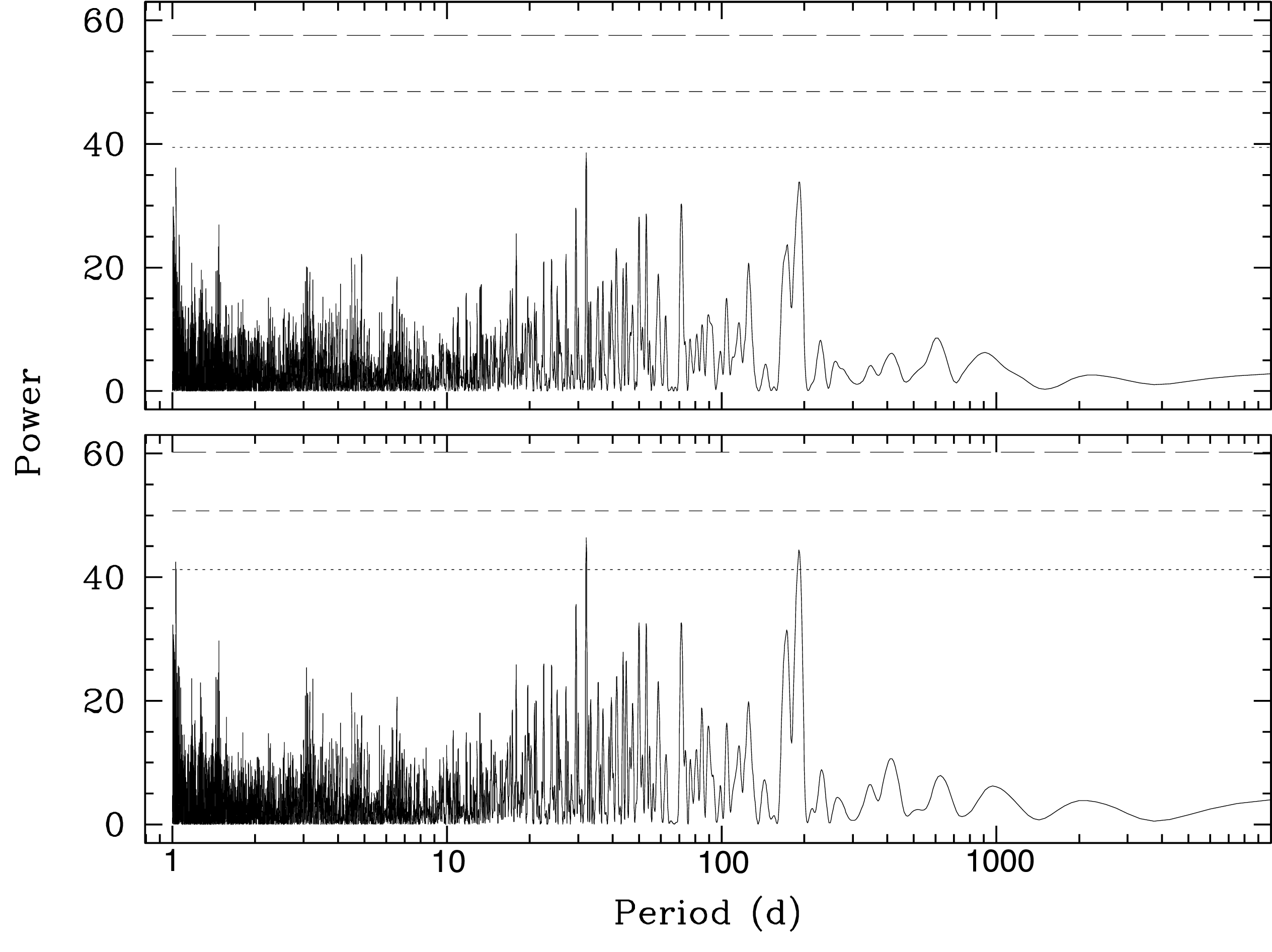}
\caption{Top panel: Periodogram of the residuals to the F11 four-planet all-circular model. Bottom panel: Periodogram of the residuals of our astrocentric circular (interacting or non-interacting) four-planet fits to all 240 HARPS RVs.}
\label{period40}
\end{figure}

\begin{table*}
\caption{Astrocentric, circular, non-interacting model for a potential five-planet system}
\label{5pl_circ_nint}
\begin{tabular}{llllll}
\hline \noalign{\smallskip}
Parameter & GJ~581\,e & GJ~581\,b & GJ~581\,c & GJ~581\,g & GJ~581\,d  \\
\noalign{\smallskip} \hline \noalign{\smallskip}
$P$ (days)                  & 3.1494$\pm$0.0305 & 5.3694$\pm$0.0122 & 12.9355$\pm$0.0591 & 32.129$\pm$0.635 & 66.671$\pm$0.948 \\
$a$ (AU)                    & 0.028459$\pm$0.000177 & 0.0406161$\pm$0.0000609 & 0.072989$\pm$0.000226 & 0.13386$\pm$0.00173 & 0.21778$\pm$0.00198 \\
$K$ (m\,s$^{-1}$)           & 1.771$\pm$0.387 & 12.76$\pm$0.94 & 3.154$\pm$0.524 & 0.985$\pm$0.282 & 2.047$\pm$0.361 \\
$m\,\sin{i}$ ($M_{\oplus}$) & 1.860$\pm$0.406 & 16.00$\pm$1.17 & 5.302$\pm$0.881 & 2.242$\pm$0.644 & 5.94$\pm$1.05 \\
MA ($^{\circ}$)              & 141.9$\pm$39.2 & 338.4$\pm$13.6 & 181.0$\pm$52.6 & 55.3$\pm$63.3 & 227.3$\pm$41.5 \\ \hline
fit epoch (JD)              & \multicolumn{5}{c}{2453152.712}\\
$N_{\rm meas}$              & \multicolumn{5}{c}{240}\\
$\chi_{\nu}^2$              & \multicolumn{5}{c}{2.570}\\
RMS (m\,s$^{-1}$)           & \multicolumn{5}{c}{1.907}\\
\noalign{\smallskip}\hline
\end{tabular}
\end{table*}

Figure~6 shows all of the phased reflex velocities from our five-planet
non-interacting circular orbit model (Table~9). Each phased curve represents the
reflex velocity of the host star caused by an individual planet with all the others subtracted off.
The curves are presented in order of increasing period: 3.15, 5.4, 12.9, 32,
and 67 days, respectively, from top to bottom. The vertical scale is held
constant for all but the 5.4-day.

\begin{figure}
\includegraphics[angle=0, width=80mm]{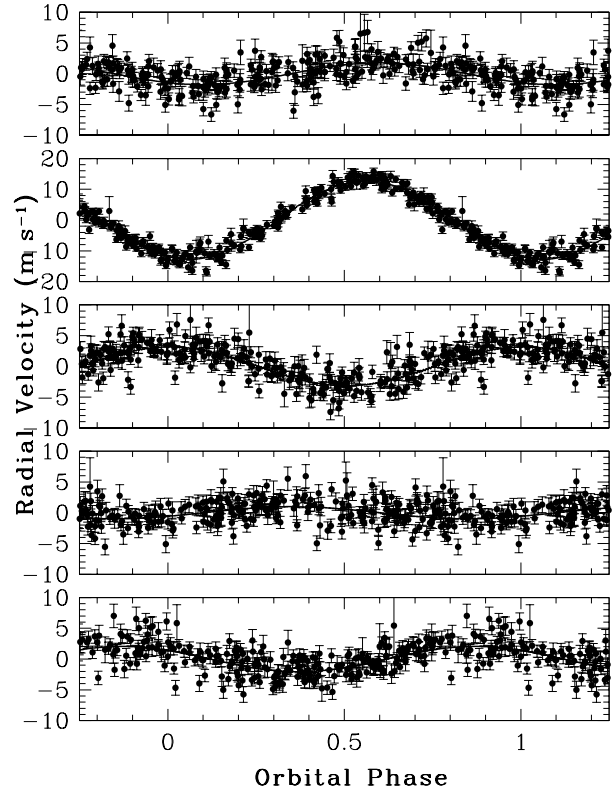}
\caption{Phased reflex velocities from the five-planet non-interacting circular orbit model. Shown successively from the top are the 3.15, 5.4, 12.9,
32, and 67-day planets. Solid lines represent the actual model.}
\label{phases}
\end{figure}

We also explored making the four-planet non-interacting circular fit fully self-consistent by including the mutual perturbations between all planets. For the sake of brevity, we refer to these models as ``circular interacting''. By this we mean that the model is a fully self-consistent N-body fit in which the osculating orbital elements at the epoch of the first RV measurement have zero eccentricity for each planet.
Table~10 shows the resulting astrocentric fit when we include these mutual
perturbations between the planets in our (initially) all-circular four-planet
fit. Turning on Gragg-Bulirsch-Stoer integration in the SYSTEMIC Console with
the four-planet non-interacting model causes the initial $\chi_{\nu}^2$ value
obtained for the zeroth iteration of the LM algorithm to jump up to $>200$, a
rather significant increase from the 2.88 value in Table~5, and demonstrates
that these circular orbits are interacting significantly over the 7-year time span covered by the HARPS
observations. Including the interactions results in rather large differences in the fitted mean anomalies and periods.
However, the final $\chi_{\nu}^2$ and RMS values indicate that
the optimized integrated and non-integrated fits are statistically identical.

The interactions between the planets in the circular-interacting model are predominantly precession-driven adjustments to the periods of the inner planets arising from the orbit-averaged axisymmetric modifications to the stellar potential generated by the planets themselves. However, just to be sure, we checked our 4-planet circular interacting model of Table 10 for long term dynamical stability using the Hybrid simplectic integrator in the Mercury integration package. We used a time step of 0.1 days and followed the system for 20 Myr. Not unexpectedly, we found the system to be extremely stable, with no significant increases in any of the eccentricities. Over the 20 Myr of the simulation, the innermost planet achieved the highest eccentricity, peaking at only 0.00285.

The 32.1-day peak that is visible in both panels of Figure~5 is also present, at the same
power, in the residuals of the integrated four-planet fit, with a similar MCFAP
of 3\%. Baluev (\cite{bal09}) presented a method for computing the upper limit to FAPS associated with signals derived from multiharmonic periodogram peaks.
We used Baluev's method as an additional check, obtaining FAP $<0.04$ for the 32-day peak. A final bootstrap algorithm run, using $10^{5}$ trials, obtained FAP $=0.037$. We did not compute the associated F-test values. However, the similarities
in the results associated with the four-planet all-circular interacting and non-interacting
models are likely to also be present in comparing the corresponding
five-planet fits.  As a result, the F-test values above for the all-circular
non-integrated five-planet fits are likely good estimates for the integrated fit.

\begin{table*}
\caption{Astrocentric, circular, interacting orbital model}
\label{4pl_circ_int}
\begin{tabular}{lllll}
\hline \noalign{\smallskip}
Parameter & GJ~581\,e & GJ~581\,b & GJ~581\,c & GJ~581\,d  \\
\noalign{\smallskip} \hline \noalign{\smallskip}
$P$ (days)                  & 3.150$\pm$0.136 & 5.3691$\pm$0.0126 & 12.9098$\pm$0.0395 & 66.60$\pm$2.61 \\
$a$ (AU)                    & 0.02846$\pm$0.00103 & 0.0406150$\pm$0.0000632 & 0.072892$\pm$0.000148 & 0.21762$\pm$0.00505 \\
$K$ (m\,s$^{-1}$)           & 1.748$\pm$0.426 & 12.747$\pm$0.906 & 3.214$\pm$0.563 & 1.810$\pm$0.367 \\
$m\,\sin{i}$ ($M_{\oplus}$) & 1.835$\pm$0.447 & 15.99$\pm$1.13 & 5.400$\pm$0.947 & 5.25$\pm$1.07 \\
MA ($^{\circ}$)              & 138.8$\pm$38.6 & 159.0$\pm$14.5 & 355.2$\pm$52.1 & 55.9$\pm$40.2 \\ \hline
fit epoch (JD)              & \multicolumn{4}{c}{2453152.712}\\
$N_{\rm meas}$              & \multicolumn{4}{c}{240}\\
$\chi_{\nu}^2$              & \multicolumn{4}{c}{2.879}\\
RMS (m\,s$^{-1}$)           & \multicolumn{4}{c}{2.010}\\
\noalign{\smallskip}\hline
\end{tabular}
\end{table*}

Figure~7 shows the periodogram of the residuals from the five-planet integrated
all-circular fit. The residuals contain no significant periodicities.
Note that the incorporation of the 32-day fifth planet to the
model also removes some of the power of the 190-day peak. However, removing the 190-day power first leaves residual power at 32 days (and 71 days). Much of the power in the 190-day peak may originate from the frequency difference between the 122.4-d and 354-d peaks in the PSW since 1/190 $\simeq$ 1/122.4 - 1/354. At the same time, there is no evidence in the present greatly expanded data set of the potential 433-day signal reported by V10, nor of the 399-day signal found in the M09 HARPS data alone set by the Bayesian analysis of Gregory (\cite{gregory11}).

\begin{figure}
\includegraphics[angle=-90, width=80mm]{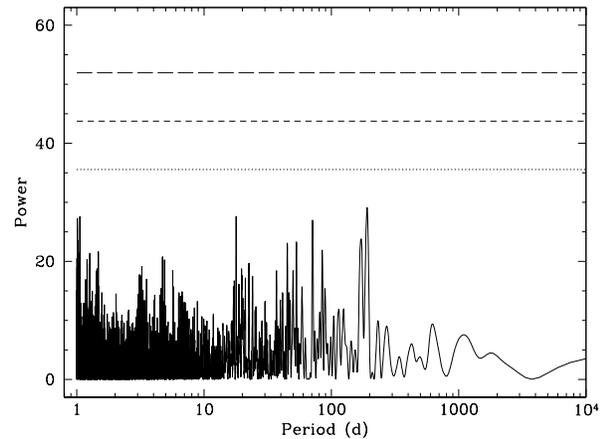}
\caption{Periodogram of the residuals of the five-planet all-circular fit.}
\label{period40_int}
\end{figure}

\section{Discussion}

Our analysis reveals that the Keplerian modeling of the GJ~581 system is considerably more complex than either of the analyses of M09 or F11 would lead to believe. In particular, allowing floating eccentricities for some or all of the components frequently leads to solutions that are dramatically (or even worse, subtly) unstable and therefore are completely unphysical, despite providing excellent fits to the data. The fitting routine used by the Geneva group is described by M09 as being ``a heuristic algorithm, which mixes standard non-linear minimizations with genetic algorithms'',  and is claimed to be able to efficiently explore the large parameter space of multi-planet systems, quickly converging on the best solution. But while the all-eccentric model of F11 may represent, according to this heuristic algorithm, some ideally-optimized overall fit, using only four planets, with no need for any further planets in the system, it is also manifestly unstable. And, as we have shown, all four planets also experience non-negligible gravitational interactions that need to be properly included in any such modeling. Indeed, the four-planet model presented by M09 was stable only because of their \emph{ad hoc} decision to hold the eccentricities of the inner two planets at zero, while allowing the outer two planets' eccentricities to float. Additionally, in their investigation of system stability as a function of inclination, M09 initially assumed zero eccentricities for the inner 3.15-d and 5.37-d planets. However, they then found that the system is even less stable at any inclination when the eccentricity of the 3.15-day planet was set to 0.1. This disagrees strongly though with their next incarnation of GJ~581, wherein F11 elected instead to allow all eccentricities to float, and in particular allowed the eccentricity of the 3.15-day to rise to 0.32, producing a better fit to the data, albeit with a highly unphysical Keplerian model.

The surprisingly high value of 0.32 for the eccentricity of the 3.15-day planet  was duly noted by F11. They described this innermost planet as ``subject to the strongest tidal forces and least expected to have high eccentricity''. That concern was, however, not considered important, and the result was held forth as their most significant eccentricity determination, at a reported significance level of 3.6\,$\sigma$. Our simulations identify this eccentricity to be the most likely contributor to system instability and show that this high of an eccentricity is completely incompatible with dynamical stability of their model. It also casts serious doubt on the reality of all their other reported eccentricities, particularly that of the 67-day planet d, which figures critically into this discussion, and for which F11 reported a lesser significance level of 2.8\,$\sigma$. There is no question that it is almost certainly possible to stabilize the F11 Keplerian model by tuning of eccentricities, and/or by simply forcing the inner planets to be circular, as was done by M09. However, such setting of eccentricities introduces biasses and personal choices into the model that inappropriately affect the resulting solution.

The relatively large uncertainties we find in Table~5 also underscore the potential pitfalls introduced by incorporating floating eccentricities into the modeling process for this system. Our 1000-trial bootstrap computation of the uncertainties reveals that acceptable solutions can often be found that vary the period of the 67-day planet by as much as $\pm$4 days. We also attempted a procedure in which we gradually fit for one, two, three, and four planets on non-interacting orbits with floating eccentricities and quickly found a dramatically different solution than the Keplerian fit published in F11. When we got up to a three-planet fit, the first two planets came in at their expected $\sim$5.4 and 12.9 days, but the third planet occasionally came in at $\sim$86 days (even though it had been started off at $\sim$67 days in the initial guess).

All these modeling forays reinforce our suspicion that the uncertainties in the parameters in Table~2 of F11 (provided also for reference in Tables 2 and 3 above) probably significantly underestimate the true uncertainties. We found that, to obtain their fit, we had to be careful in deciding which parameters to temporarily hold fixed while allowing others to float. A full-on analysis of this type, and an analysis to attempt to find a stable Keplerian (or more likely Newtonian) model with all-floating eccentricities would be worthwhile. We have begun such a study, however this is well beyond the scope of the present paper.

As mentioned in the introduction, the particular case of GJ~581 is further complicated by the connection between (1) the adopted eccentricity of the 67-day planet GJ~581d,  (2) potential planets near half that period, and (3) sampling aliases. These complications are described in detail by AD11. Basically, eccentricity harmonics of a known planet can sometimes mask the signal of other planets near half of that planet's period. Any fitting sequence for the GJ~581 system that proceeds sequentially in order of signal strength (as all previous modelers, including the Bayesian studies have done), will necessarily fit the 67-day planet ahead of any potential fifth planet in the system. If the modeler elects to allow the eccentricity of the 67-day planet to float, least-squares fitting routines will take advantage of this extra degree of freedom, allowing the eccentricity of the 67-day to rise, and thereby largely masking any signal from a real fifth planet near half that period. Aliases from the unevenly-spaced sampling in the data set further complicate the behavior of peaks at or around half the period of the 67-day. Despite these potential complications, AD11 used Monte Carlo simulations of the effects of both the eccentricity harmonic and its aliases to conclude that the presence of GJ~581g was well-supported by the data set analyzed in V10.

Since the 67-day signal is not far from twice the lunar synodic period, there was a distinct but subtle phase gap (or phase paucity) present in the M09 data set that could easily trigger an unduly eccentric solution. Over-usage of eccentricity can look attractive from a least-squares standpoint if it avoids incurring a $\chi_{\nu}^2$ penalty by phasing its largest residuals to fall in that phase gap. In V10, the propensity of that subtle ``phase paucity'' in the M09 data to trigger eccentric fits for the 67-day planet was discussed. V10 also pointed out that the HARPS and HIRES data sets just did not merge well under the assumption of all-floating-eccentricity fits, leaving larger numbers of peaks in the residuals periodograms. By contrast, models that assumed all-circular orbits allowed the two data sets to meld much more closely, and produced equivalent quality fits with fewer parameters. Thus they were formally superior in a strict $\chi_{\nu}^2$ sense. Allowing the eccentricities of all four known planets to float adds 8 additional parameters to the model, more additional degrees of freedom than adding even two more planets on circular orbits. The principle of parsimony clearly favors an all circular model.

V10 thus found, as we also conclude here, that all-circular-orbit models for GJ~581 are much more compelling and well-founded than using floating eccentricities, and fit the data as well or better with fewer parameters. This approach also respects the fact that most of the planet signals in the GJ~581 data set are near or even below the noise level set by the unknown stellar jitter and by remaining unknown systematics in the RV reduction pipelines. It does not seem justified to us to presume the ability to discern the shapes of such weak signals by invoking two extra parameters (eccentricity and longitude of periastron) for each planet. Or, put another way, if all eccentricities are consistent with zero within their formal uncertainties, allowing non-zero eccentricities unnecessarily invites over-fitting the noise and/or phase gaps, not a judicious modeling approach.

The result from dynamical studies, that F11's allegedly most significant eccentricity detection (for GJ~581e) must rather be at or nearly circular, raises legitimate skepticism about the significance of all of the other eccentricity values reported with substantially less significance in their Keplerian model, especially the reported eccentricity of the 67-day planet d that could be masking a planet near half that period. Our modeling studies suggest that the present HARPS data set of F11 offers no evidence for significant eccentricity for any of the four known planets. There have also been two Bayesian studies that each lend support to the all-circular approach. Tuomi (\cite{tuomi11}) and  Gregory (\cite{gregory11}) both carried out Bayesian analyses of the full [HARPS + HIRES] data set analyzed by V10. Using the full combined data set, neither Bayesian study found evidence of more than four planets in this system, though Gregory (\cite{gregory11}) did indeed find support, from the HARPS data alone, for a 5th planet at 399 days (presumably identifiable with GJ~581f). At the same time, the Tuomi (\cite{tuomi11}) study explicitly concluded that the orbits of the four confirmed planets were all consistent with circular. Tuomi (\cite{tuomi11}) cited 99\% Bayesian credibility ranges of [0-0.43] for the eccentricity of the 3.15-day planet, [0-0.05] for the 5.4-day planet, [0-0.29] for the 12.9-day planet, and [0-0.67] for the 67-day planet. The Bayesian analysis of Gregory (\cite{gregory11}) also lists uncertainties for 3 of 4 eccentricities in this system that are consistent with circular.

These Bayesian studies are, however, not without their own problems, nor are they above criticism. Neither Bayesian study discussed nor referenced the eccentricity harmonic issue raised by AD11. And since both Bayesian studies modeled the system in descending order of signal amplitude, they encountered the 67-day signal first and would have allowed its eccentricity to rise, thereby becoming blind to planet g hidden in the 67-day orbit's eccentricity harmonic. Bayesian analyses are well-known to suffer from a propensity to over-use eccentricity and it is thus not unexpected that they would fall easy victim to this eccentricity harmonic pitfall. It would be interesting to re-do the Bayesian analyses using priors that discourage use of eccentricity and hold all the orbits nearly circular.

Perhaps most seriously, neither Bayesian study included dynamical stability criteria in their analyses since they do not include planet-planet gravitational interactions. Both Bayesian studies treated the orbits as simple summed Keplerians, an approach that is demonstrably inadequate for this system given the observed magnitude of these gravitational interactions over the time spans of the data sets. Nor did either Bayesian analysis include any accounting for dynamical instability. As a result, Bayesian analyses of systems having eccentric orbits are colored by a large, entirely uncontrolled admixture of demonstrably unstable cases.

The fact that neither Bayesian analysis found sufficient evidence for more than four planets in the system also deserves further scrutiny. Tuomi (\cite{tuomi11}) adopted the traditional Bayesian evidence ratio threshold of 148:1 for a convincing detection. However, Jenkins \& Peacock (\cite{jenkins11})  have more recently raised serious caveats about the choice of this traditional threshold for the Bayesian Evidence Ratio. They conclude that the traditional assumption of a Bayesian evidence ratio (or Bayes factor) of 148:1 is excessively conservative, the equivalent of a 5.5\,$\sigma$ threshold. Jenkins \& Peacock (\cite{jenkins11}) warn that ``setting the critical odds at the apparently desirable 148 to 1 means we will rarely exceed the evidence ratio threshold. As with any classical test statistic, it makes no sense to set a critical value which will hardly ever be exceeded for the amount of data available''.

The simple fact is that the signal levels for any and all planets beyond the first four in the GJ~581 system do not yet rise to this very conservative 5.5\,$\sigma$ level of significance. So, contrary to the widespread impression that the Bayesian results rule out any more than 4 planets in the GJ~581 system, the additional planet claims of V10 are actually not in discord with these Bayesian analyses. Using such an excessively conservative  threshold, neither Bayesian analysis should have been able to confirm either of the V10 claims since those signals are well below a 5.5\,$\sigma$ threshold, at significance levels arguably no higher than 4-5$\sigma$. Additionally and even more fundamentally, Jenkins \& Peacock (\cite{jenkins11}) found the Bayesian evidence ratio to be a noisy statistic, and cautioned that it may not be sensible to accept or reject a model based solely on whether that evidence ratio reaches some threshold value. They conclude that the performance of such Bayesian tests is significantly affected by the signal to noise ratio in the data, as well as by the assumed priors, and by the particular threshold in the evidence ratio that is taken as decisive.

Finally, there is the recent detailed and thorough re-analysis of the M09 and V10 data sets by TDS12. In agreement with AD11, they also find that the existence of planet g is intimately connected to the eccentricity of the orbit of the 67-day planet d, and it is not possible to disconnect the existence of the former planet from the determination of eccentricity of the latter planet. They do find evidence for a signal at 450 days, near the period of 433 days reported by V10 for GJ 581f. However this signal is too near the detection limit of their analysis, and in a region heavily confounded by aliases.

Assuming circular orbits, the TDS12 analysis found minima in the RMS of their fits for two favored periods for GJ 581g, one near $\sim 33$ days, similar to that presented in Table 9 above from our re-analysis of the F11 data set, and the other near $\sim 36$ days, close to the value of 36.6 days claimed for GJ 581g by V10. It is not clear which, if either, of these two choices might be the true period for planet g, and which might be a yearly alias of that true period. As TDS12 point out, the difference between these two minima in the solution RMS is caused by an alias indetermination from observations concentrated near the oppositions. The true period of planet g might be either of these values, with the other peak being the yearly alias of that true period. Udry et al. (\cite{udry07})  were similarly confused in their original determination of the period of planet d. Their originally reported period of 82 days turned out to be the yearly alias of the true period of 67 days, as 1/82 $\sim$ (1/67 - 1/365.25). This was subsequently corrected in M09 to the true period of 67 days for planet d. Similarly, the 36-day signal reported by V10 for GJ 581g could well have been the yearly alias of a true period of 33 days, as 1/33 $\sim$ (1/36 + 1/365.25). Perhaps the greatly expanded data set of F11 has now resolved this ambiguity in favor of the 33-day period. Whichever is the case, within the limits of aliasing effects in the present data set, both the 33-day and 36-day signals are mutually consistent with a 5th planet in the system at one or the other period.

\section{Conclusion}

We have carried out an extensive re-analysis of the full 240-point HARPS precision RV set for GJ~581 presented by F11, as well as a re-assessment of their analyses of these data. We explored a wide range of models with both non-interacting and interacting orbits. Our analysis leads us to conclude that the $\chi_{\nu}^2$ and RMS values reported by F11 reveal that some of the worst-fitting data points to their model were apparently omitted from their analysis, thereby specifically suppressing observational evidence for any further planets in the system beyond those four claimed in their model. We also carried out a suite of 4000 N-body simulations of the F11 Keplerian model. Not one of these 4000 trials remained stable for more than 200,000 years. This result shows that the F11 Keplerian model is extremely unstable and is therefore manifestly untenable. All unstable orbits ended in the merger of the 3.15-day and 5.4-day planets. The main destabilizing factor is F11's relatively high value (0.32) for the eccentricity of the 3.15-day innermost planet. Such a high value is completely incompatible with system stability, and is also unexpected from tidal circularization considerations. The marked lack of stability underscores the potential pitfalls of incorporating floating eccentricities into such modeling and makes all-circular models more compelling and well-founded for such systems (systems with multiple extremely low-amplitude signals closely packed in period space).

Based on their four-planet non-interacting Keplerian fit to the HARPS data, F11 concluded that the present 240-point HARPS data set, a factor of two larger now than that of M09, contains no evidence for any planets beyond the four already announced by M09 and confirmed by V10. But we have shown in the present work that the F11 Keplerian solution is dramatically unstable over a wide range of starting conditions, and is thus untenable. F11's conclusion of there being only four planets in the system was based on this unphysical model and can thus be discounted. Furthermore, the data points that were apparently omitted from the F11 analysis were dropped solely based on deviation from their 4-planet model, thus unfairly and specifically suppressing evidence for any additional planets in the system. At the same time, F11 did present a viable stable four-planet all-circular model, though they did not present its residuals periodogram or any discussion of the residuals to their all-circular fit.

We developed our own four-planet all-circular models (both with and without dynamical interactions) that closely mirror the four-planet all-circular non-interacting model of F11. Contrary to F11's conclusions, we find that the full 240-point HARPS data set, when properly modeled with self-consistent stable orbits, by and of itself actually offers confirmative support for a fifth periodic signal in this system near 32-33 days, and is consistent with the possibility of having been detected as GJ 581g at its 36-day yearly alias period by V10. The residuals periodograms both of our interacting and non-interacting fits and of the F11 four-planet circular fit reveal distinct peaks near 32 days and 190 days. Both of these residuals peaks are largely simultaneously accounted for by adding a fifth planet at 32.1 days to the system. Under the assumption, now strongly supported by two Bayesian studies, that the first four planets are in circular or nearly circular orbits, this 32-day residuals signal has an empirically-determined Monte Carlo false alarm probability of 3.7\% and a Baluev-style FAP upper limit of 4\%. It is consistent with a fifth planet of minimum mass 2.2 \mearth\ in the system, orbiting at 0.13 AU, solidly in the star's classical liquid water Habitable Zone.

It may prove exceedingly difficult to break the degeneracy between the existence of a 32-day planet g and the presence of eccentricity in the orbit of planet d. The principle of parsimony and dynamic stability clearly favor an all-circular-orbits 5-planet model over an all-eccentric 4-planet model. The all-circular-orbits model is further supported by both existing Bayesian studies. Only further data and time may provide the answer. But with the 240 HARPS velocities from F11, plus another 122 HIRES velocities from V10 already in hand, it will be hard, as already noted by F11, to make further major gains in sensitivity through gains in the square root of N. Combining data sets to improve the square root of N may also introduce subtle systematics that might further confound the situation. Nevertheless, over the past year, we have continued to observe GJ 581, obtaining another observing season's worth of Keck and Magellan PFS RVs. We are also making further improvements to our data reduction pipelines with the goal of eventually compiling a data set sufficient to lift this degeneracy.

\acknowledgements
The authors would like to thank Drs. Eugenio Rivera and Greg Laughlin for many useful discussions and assistance with dynamical calculations.
SSV gratefully acknowledges support from NSF grant AST-0307493. RPB gratefully acknowledges support from NASA OSS Grant NNX07AR40G, the NASA Keck
PI program, and from the Carnegie Institution of Washington. NH acknowledges
support from the NASA Astrobiology Institute under Cooperative Agreement
NNA09DA77A at the Institute for Astronomy, University of Hawaii, and NASA
EXOB grant NNX09AN05G. The work herein is based on observations obtained
at the W. M. Keck Observatory, which is operated jointly by the University of California
and the California Institute of Technology, and we thank the UC-Keck, UH, and NASA-Keck
Time Assignment Committees for their support. We also wish to extend our special thanks
to those of Hawaiian ancestry on whose sacred mountain of Mauna Kea we are privileged
to be guests. Without their generous hospitality, the Keck observations presented herein
would not have been possible.

\end{document}